\documentclass[final]{svjour3}
\usepackage{graphicx}
\usepackage{rotating}
\usepackage{amssymb}
\usepackage{mathptmx}
\usepackage[super]{natbib}

\makeatletter
\journalname{Journal of Low Temperature Physics}
\bibpunct{}{}{,}{s}{}{,}

\begin{document}

\title{DESHIMA~2.0: Rapid redshift surveys and multi-line spectroscopy of dusty galaxies}

\author{M. Rybak$^{1,2,\dagger}$ \and
T. Bakx$^{3,4}$ \and J. Baselmans$^{1,5}$ 
\and \\ K. Karatsu$^{1,5}$ \and K. Kohno$^{6,7}$ \and T. Takekoshi$^{8,6}$ \and \\ Y. Tamura$^{3}$ \and A.
Taniguchi$^{3}$ \and P. van der Werf$^{2}$ \and \\ A. Endo$^{1}$
}

\institute{
\at $^\dagger$ \email{m.rybak@tudelft.nl} 
\at $^1$ Faculty of Electrical Engineering, Mathematics and Computer Science, Delft University of Technology, Mekelweg 4, 2628 CD Delft, the Netherlands
\at $^2$ Leiden Observatory, Leiden University, Niels Bohrweg~2, 2333~CA Leiden, the Netherlands
\at $^3$ Division of Particle and Astrophysical Science, Graduate School of Science, Nagoya University, Furocho,
Chikusa-ku, Nagoya, Aichi 464-8602, Japan
\at $^4$ National Astronomical Observatory of Japan, 2-21-1 Osawa, Mitaka, Tokyo 181-8588, Japan
\at $^5$ SRON -- Netherlands Institute for Space Research, Niels Bohrweg 4, 2333 CA Leiden, the Netherlands 
\at $^6$ Institute of Astronomy, Graduate School of Science, The University of Tokyo, 2-21-1 Osawa, Mitaka, Tokyo 181-0015, Japan
\at $^7$ Research Center for the Early Universe, Graduate School of Science, The University of Tokyo, 7-3-1 Hongo, Bunkyo-ku, Tokyo 113-0033, Japan
\at $^8$ Kitami Institute of Technology, 165 Koen-cho, Kitami, Hokkaido 090-8507, Japan
}

\authorrunning{M. Rybak \& DESHIMA collaboration} 

\maketitle

\begin{abstract}

We present a feasibility study for the high-redshift galaxy part of the Science Verification Campaign with the 220-440 GHz \textsc{Deshima} 2.0 integrated superconducting spectrometer on the ASTE telescope. The first version of the \textsc{Deshima}~2.0 chip has been recently manufactured and tested in the lab. Based on these realistic performance measurements, we evaluate potential target samples and prospects for detecting the [CII] and CO emission lines. The planned observations comprise two distinct, but complementary objectives: (1) acquiring spectroscopic redshifts for dusty galaxies selected in far-infrared/mm-wave continuum surveys; (2) multi-line observations to infer physical conditions in dusty galaxies. 


\end{abstract}

\section{Introduction}
\label{intro}

Throughout cosmic history, more than half of all the stars form in dust-obscured galaxies\cite{Madau2014, Casey2014, Zavala2021}. Due to their massive dust reservoirs, these dusty star-forming galaxies (DSFGs) are often invisible in the optical / near-IR part of the spectrum but bright in the far-infrared (FIR) to (sub)-mm wavelengths. Consequently, thousands of DSFGs have been identified in wide-field continuum surveys in the 0.1-2.0~mm regime\cite{Casey2014, Geach2016, Zavala2021}. 

However, studies of DSFGs suffer from a considerable \emph{redshift bottleneck}. This is because long-wavelength continuum observations provide only weak constraints on the redshift of individual galaxies - a critical prerequisite for further emission line studies and high-resolution imaging with interferometers such as ALMA\cite{Hodge2020}. Similarly, the optical and near-IR spectroscopic follow-up is often inefficient due to the high extinction in DSFGs, particularly in the FIR-brightest sources\cite{Casey2014}. 

Consequently, (sub)mm-wave spectroscopy has become the key to obtaining robust spectroscopic redshifts for dusty galaxies at high redshift. This is chiefly through the rotational emission line of $^{12}$CO and the fine-structure transition of C$^+$, the 158-$\mu$m [CII] line (rest-frame frequency $f_0$=1900.5~GHz). Typically, these are conducted using heterodyne receivers via spectral scans, requiring multiple instrument tunings. 

Alternatively, several dedicated wide-band instruments have been developed: e.g., grating spectrometers such as the now-defunct Z-Spec\cite{Naylor2003,Bradford2004} at the Caltech Submilimeter Observatory (CSO; 190-305~GHz) or wideband heterodyne receivers such as the ZSpectrometer\cite{Harris2007} at the Green Bank Telescope (26.5-40~GHz) and the Redshift Search Receiver (RSR\cite{Erickson2007}, 73-111~GHz) on the Large Millimeter Telescope. While these have allowed redshift measurements out to $z\simeq6$\cite{Zavala2018}, mainly using the CO emission lines, due to the relative faintness of these lines, such observations are limited to the rare, very bright galaxies. Critically, the bright [CII] fine-structure line - ideal for rapid redshift measurements due to its brightness - is generally beyond the reach of these instruments. 

To properly exploit the [CII] line for redshift measurements, wideband spectroscopy must be extended to higher frequencies, i.e. the 350-GHz and 400-GHz atmospheric windows. These frequency bands are particularly promising for spectroscopic confirmation of DSFGs, because the number density of DSFGs (and thus [CII] emitters) peaks between $z=2-4$\cite{Popping2016}, corresponding to [CII] being redshifted to 380-600~GHz.

In addition to measuring redshifts, mm-wave spectroscopy provides critical insights into the physical conditions in DSFGs. Namely, observations of multiple chemical species (e.g., CO, C$^+$, C, O, dust continuum) can be linked to the underlying physical conditions (e.g., gas density and temperature, irradiation, turbulence) using chemical and radiative transfer modelling. 
For example, the $J_\mathrm{upp}\geq8$ CO lines are sensitive to non-thermal gas excitation, such as heating by X-rays and cosmic rays or turbulence, which might be significant in intensely star-forming DSFGs. Indeed, recent studies point towards highly excited CO rotational lines in strongly lensed DSFGs\cite{Harrington2021,Riechers2021}. The octave-wide bandwidth offered by the ISS architecture is ideally suited to simultaneous observations of multiple high-excitation CO lines in high-redshift galaxies.

\section{DESHIMA~2.0: instrument description}
\label{sec:deshima}

\textsc{Deshima} (DEep Spectroscopic HIgh-redshift MApper) is an integrated superconducting spectrometer (ISS) operating in the mm-wave regime\cite{Endo2019a, Endo2019b}, combining on a single chip a superconducting filterbank with an array of microwave kinetic inductance detectors (KIDs). Several other KID mm-wave spectrometers are currently under development, e.g., CONCERTO\cite{Monfardini2021, Catalano2021} (130--310~GHz), and Super-Spec\cite{Shirokoff2012, Karkare2020} (190--310~GHz).

The first version of the instrument - \textsc{Deshima}~1.0 - achieved the first light in 2017\cite{Endo2019b, Takekoshi2020}. 
These demonstrated the spectroscopy of point sources (post-AGB star IRC+10216, merging galaxy pair VV~114 at redshift $z=0.02$) and the on-the-fly spectroscopic mapping of extended regions (the Orion KL star-forming region and the nearby galaxy NGC~253).

In 2022, an upgraded \textsc{Deshima} 2.0 spectrometer will be installed at the ASTE (Atacama Submillimeter Telescope Experiment) 10-meter telescope\cite{Ezawa2004} in the Atacama desert, Chile, at an altitude of 4,860~metres. \textsc{Deshima} 2.0 will provide an instantaneous frequency coverage of 220 - 440~GHz at $R\simeq500$ ($\Delta v \simeq$ 600 km/s). 
Besides the significantly expanded bandwidth, major upgrades between \textsc{Deshima}~1.0 and \textsc{Deshima}~2.0 include a leaky-lens antenna\cite{Hahnle2020}, improved filter design, and a sky-position chopper\footnote{For a detailed summary, see the contribution by Taniguchi et al., in this volume\cite{Taniguchi2021}.}. Together, these upgrades result in a factor 4-8 improvement in sensitivity over \textsc{Deshima} 1.0. Further sensitivity improvements might be achieved explicitly modelling the instrument and atmospheric noise, rather than simply subtracting the on- and off-source spectra\cite{Taniguchi2021a}.

The first version of the \textsc{Deshima}~2.0 on-chip filterbank has been recently manufactured and tested in the lab\cite{Taniguchi2021}. The filters cover almost the entire target bandwidth, with a mean peak coupling efficiency of 14\%, increasing up to 30-50\% for some channels (target: $\sim$30
\%). The main source of discrepancy between the current and target performance is uneven channel spacing, which reduces the coupling efficiency of individual channels. 

\begin{figure*}
\begin{center}
  \includegraphics[width=0.9\textwidth]{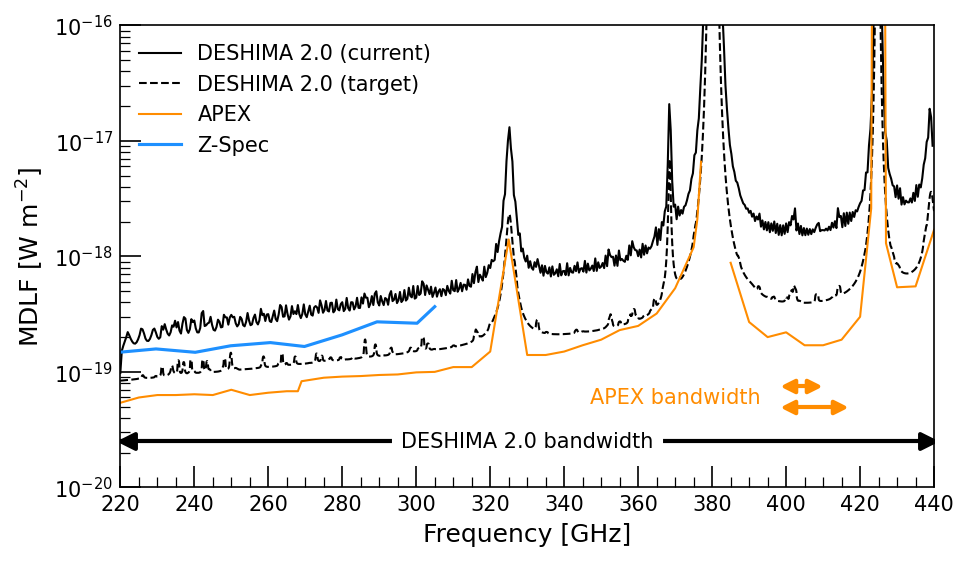}
\caption{5$\sigma$ minimum detectable line flux (MDLF) for a 600 km/s wide line for the current and target \textsc{Deshima}~2.0 performance\cite{Taniguchi2021}, compared to the Z-Spec\cite{Inami2008} and the current suite of APEX receivers. For all instruments, we assume a precipitable water vapour of 0.6~mm, target elevation of 60~deg and on-source time of 3.6~hr. The instantaneous bandwidth of \textsc{Deshima} 2.0 corresponds to 20 APEX tunings (APEX bandwidth = 8.0-15.8~GHz, depending on the receiver). }
\label{fig:deshima_mdlf}  
\end{center}
\end{figure*}

In Fig.~\ref{fig:deshima_mdlf}, we show the current and target \textsc{Deshima}~2.0 performance compared to Z-Spec\cite{Inami2008} at the CSO on Mauna Kea, and the current suite of receivers on the 12-m Atacama Pathfinder EXperiment (APEX) telescope (nFLASH230, SEPIA345, nFLASH460\footnote{The APEX sensitivity calculations are based on \texttt{https://www.apex-telescope.org/ heterodyne/calculator/ns/index.php}}), which is just a few kilometres away from ASTE. Compared to the Z-Spec, \textsc{Deshima}~2.0 has sensitivity 1.5--2.0$\times$ lower for a given pwv value. However, ASTE has more favourable weather conditions (pwv=0.6~mm corresponds to the top 25th percentile for ASTE, but only the 10th percentile for Mauna Kea\cite{Garcia2010}). \textsc{Deshima}~2.0 is thus competitive with Z-Spec, with the added advantage of covering the 305-440~GHz range. Compared to APEX, the current \textsc{Deshima}~2.0 sensitivity is 4--5$\times$ lower; the target sensitivity will be $\sim$2$\times$ higher. At that point, science objectives that would require four or more APEX tunings will be more economically achieved with \textsc{Deshima}~2.0. For such applications, \textsc{Deshima}~2.0 will be directly competitive with APEX.

\section{DESHIMA~2.0: Science Verification Campaign targets}
\label{sec:svc}

The primary targets for \textsc{Deshima} are DSFGs with large apparent FIR luminosity (i.e. $L_\mathrm{8-1100\mu m}$ of few $\times10^{13} L_\odot$), many of which are strongly gravitationally lensed. In fact, hundreds of strongly lensed dusty galaxies were discovered in wide-field continuum surveys at FIR and sub-mm wavelengths with \textit{Herschel}\cite{Negrello2010, Negrello2017, Wardlow2013} and \textit{Planck} satellites\cite{Canameras2015} and the South Pole Telescope\cite{Vieira2013} (SPT).

\begin{figure}
\begin{center}
  \includegraphics[width=0.95\textwidth]{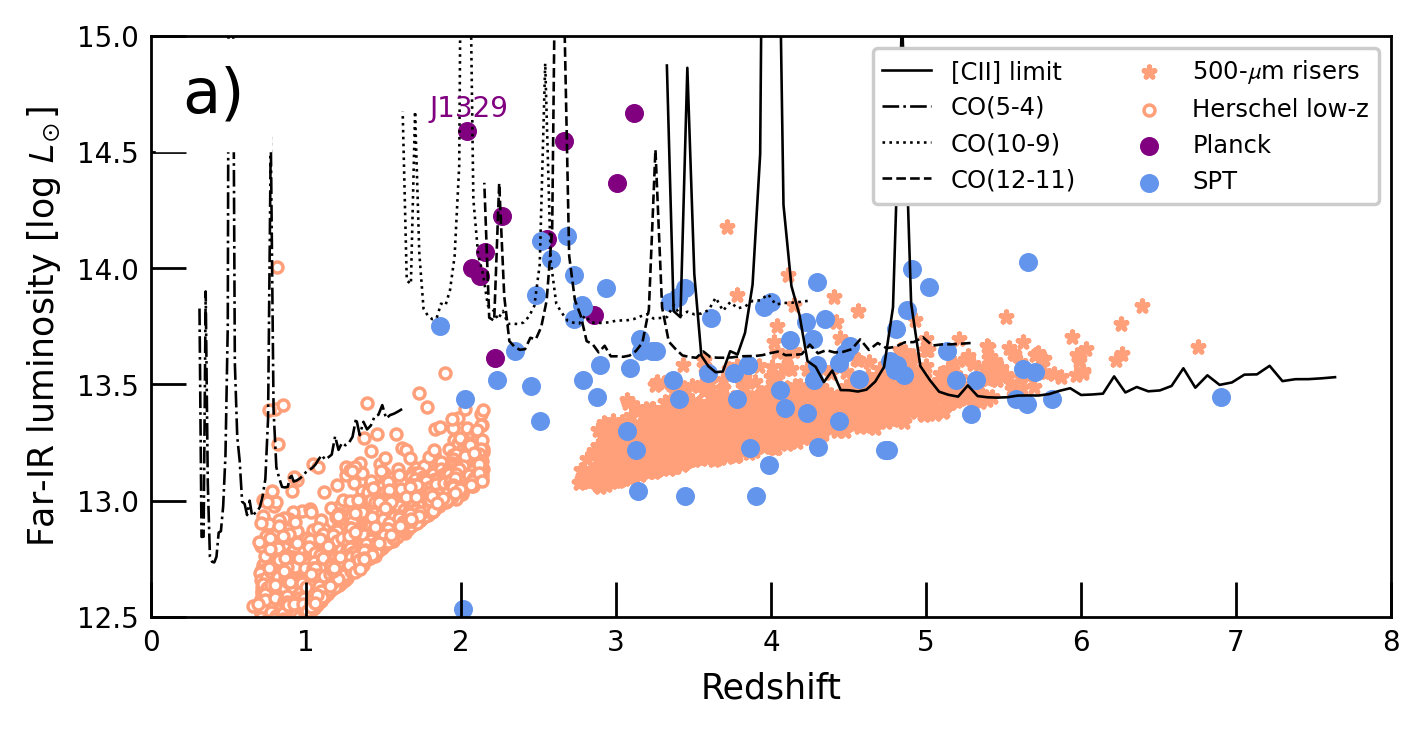}\\
    \includegraphics[width=0.95\textwidth]{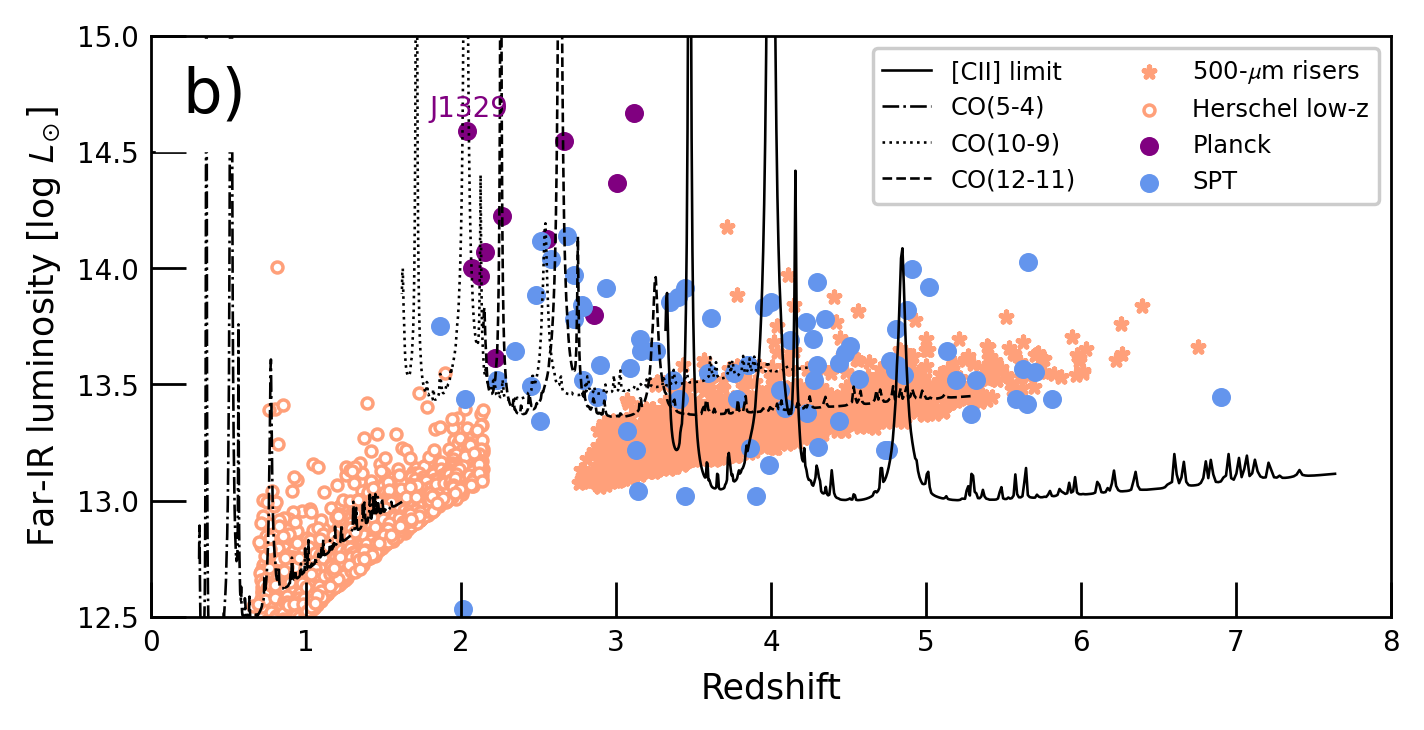}
\caption{Redshift distribution and apparent FIR luminosity of bright DSFGs observable from the ASTE site from the SPT, \textit{Planck} and the \textit{Herschel} DSFGs, compared to \textbf{a)} the current and \textbf{b)} target\textsc{Deshima}~2.0 performance (\textit{lower}). The black lines indicate the 5$\sigma$ detection limits for the CO(5--4), (10--9), (13--12), and [CII] emission from individual sources, assuming empirical CO-FIR ratios\cite{Kamenetzky2016,Greve2014} and a [CII]/FIR ratio of $10^{-3}$. We assume 12-hr observations at precipitable water vapour of 1.0 mm, i.e. less than two median observing nights. Galaxies at $z\leq3.3$ have the [CII] line outside \textsc{Deshima} 2.0 frequency coverage, but will be prime targets for multi-line spectroscopy. We highlight J1329+2243 - a very bright $z=2.05$ lensed DSFGs, a simulated spectrum for which is presented in Fig.~\ref{fig:j1329}. The current \textsc{Deshima}~2.0 performance already allows science-grade observations of high-redshift galaxies.}
\label{fig:targets_z_LFIR}  
\end{center}
\end{figure}

Fig.\,\ref{fig:targets_z_LFIR} shows the redshift and apparent FIR luminosity distribution of DSFGs from the SPT and \textit{Planck} samples (virtually all with spectroscopic redshifts) and the \textit{Herschel}-selected high/low-redshift samples.
For comparison, we show the limiting FIR luminosity of sources for which the CO(5--4), (10--9), (13--12) and the [CII] line can be detected at 5$\sigma$ level in 5-hr on source ($\sim$12 hr total with overheads). These are based on empirical CO--FIR relations from Kamenetzky et al.\cite{Kamenetzky2016} for CO(5--4), Greve et al.\cite{Greve2014} for CO(10--9) and (13--12), and $L_\mathrm{[CII]}/L_\mathrm{FIR}$ ratio of $10^{-3}$, typical for DSFGs\cite{Gullberg2015, Rybak2019}. The CO (and [CII]) luminosities in individual galaxies can deviate from these trends by up to 1~dex\cite{Greve2014, Mashian2015,Kamenetzky2016, Harrington2021}; large-sample surveys with \textsc{Deshima}~2.0 will further constrain the range of CO excitation in DSFGs.

\subsection{Spectroscopic redshifts for bright \textit{Herschel}-selected galaxies}

One of the primary aims of the \textsc{Deshima}~2.0 Science Verification Campaign is to demonstrate the rapid redshift acquisition capability. The atmospheric windows and bandwidth of \textsc{Deshima} offer two promising regions for efficient redshift searches. 

Firstly, at lower-redshift ($z \sim 0.5 - 2.0$), the large bandwidth of \textsc{Deshima} offers multiple-line detections of galaxies. Counterintuitively, finding robust redshifts of lower-redshift dust-obscured galaxies is challenging. While the spectral lines suffer less from the cosmological dimming, wider bandwidths are necessary to cover the entire possible redshift space. In addition, DSFGs at $z\leq2$ are undetected in SPT and \textit{Planck} survey, while most \textit{Herschel} redshift follow-up prioritises high-$z$ targets. The \textit{Herschel} catalogues have thus left $\sim 4000$ $z \sim 1$ targets  unexplored. As $z\leq1$ DSFGs have low lensing probability \citep{weiss2013}, this sample presents a population of intrinsically-bright galaxies after the peak of the cosmic star-forming activity \citep{Madau2014}, i.e. when galaxy-wide quenching should be in full effect. \textsc{Deshima}’s 220-GHz bandwidth will allow fast redshift acquisition for these sources, removing degeneracies in redshift due to the wide spacing of the CO lines at low redshift \citep{Bakx2020IRAM}. 

As for the high-redshift end, while most bright DSFGs have secure spectroscopic redshifts, a large population of DSFGs with lower apparent luminosities remains unexplored. Namely, the $\geq$1000~deg$^2$ \textit{Herschel} footprint\cite{Shirley2019,Shirley2021} provides a sample of $\approx$ 2000 ``500-$\mu$m risers'': DSFGs with flux density peaking at/beyond 500-$\mu$m. A 500-$\mu$m rising colour selection, relative to 250 and 350 $\mu$m, promises to select the highest-redshift \textit{Herschel} candidates\cite{ivison16}, with $z_{\rm phot} \geq 3.5$. A major advantage of \textsc{Deshima}~2.0 is the wideband spectroscopy in the 385-440 GHz band (interrupted by the 425-GHz telluric line), corresponding to the lower half of ALMA Band~8. This enables [CII] observations at $z=3.3-3.9$, the epoch when the [CII] luminosity function is predicted to peak\cite{Popping2016}. 

We expect to invest a total of 400~hr for both the low- and high-redshift goal (200~hr per goal); this should yield robust redshifts for $\sim 15 - 20$ galaxies each.

\begin{figure}
\begin{center}
  \includegraphics[width=0.99\textwidth]{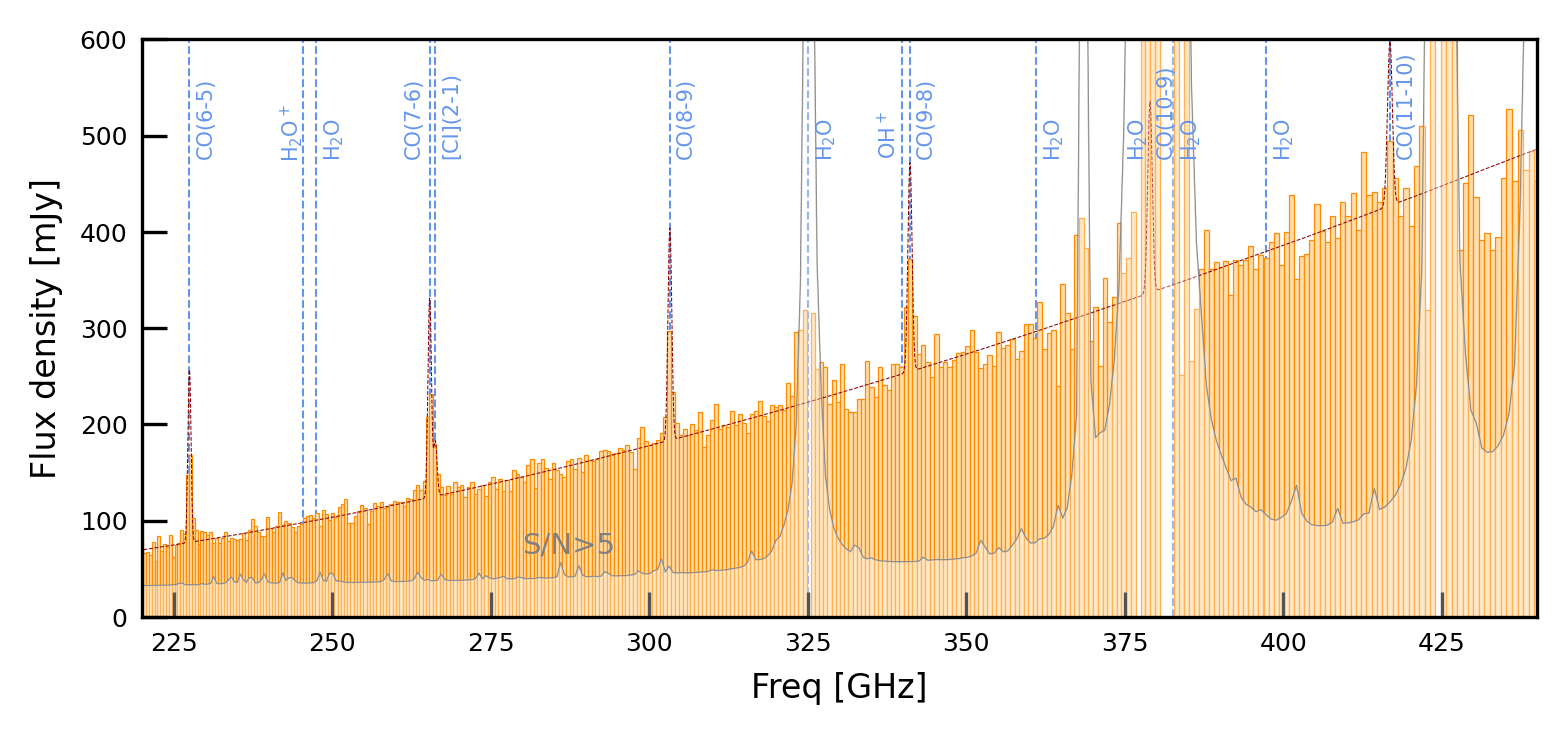}
\caption{A simulated spectrum of J1329+2243, assuming 3 hours on-source at source elevation of 40~deg and pwv=0.6~mm. This very bright source is the best target for CO ladder observations with \textsc{Deshima}~2.0. The red line shows the input spectrum based on existing observations\cite{Harrington2021}. The grey line denotes the S/N=5 threshold for each channel; increasing the on-source time will move the grey line downwards. We expect to detect four CO rotational lines; our observations will also cover several other potentially bright lines - a key discovery space for \textsc{Deshima}~2.0.}
\label{fig:j1329}  
\end{center}
\end{figure}

\subsection{Multi-line spectroscopy of bright lensed DSFGs}

Because of their high apparent brightness, the lensed DSFGs selected by \textit{Planck} and SPT are ideally suited for demonstrating multi-line spectroscopy capabilities of \textsc{Deshima}~2.0. Multi-line spectroscopy is critical for constraining different physical properties. For example, the [CII] 158-$\mu$m line is a sensitive probe of the far-UV irradiation. On the other hand, the excitation of CO rotational lines is primarily driven by gas density\cite{Stacey1991,HollenbachTielens1999}. Depending on the complexity of the data, the models can range from static, one-dimensional gas slabs\cite{Kaufman1999, Kaufman2006, VanDerTak2007, Pound2008} to fully three-dimensional models\cite{Bisbas2012} and might incorporate time evolution.

Previous studies with wide-band spectrometers such as Z-Spec have been limited to a handful of very bright sources\cite{Bradford2009, Bradford2011, Lupu2012}; with \textsc{Deshima}~2.0, we will expand this approach to a much larger sample of DSFGs. In particular, apart from the very bright \textit{Planck}-selected sources, the high-excitation CO emission in DSFGs remains almost completely unexplored. \textsc{Deshima}~2.0 should detect the high-$J$ CO lines in the brightest \textit{Planck} and SPT DSFGs in only a few hours on-source.

As a demonstration of a pilot observation, Fig.~\ref{fig:j1329} shows a simulated spectrum of J1329+ 2243. J1329+2243 is the most FIR-luminous source at $z\geq3$ from the samples considered in Fig.~\ref{fig:targets_z_LFIR}, with extensive archival CO observations\cite{Harrington2021}. As for the chip properties, we adopt a nominal R=500 design with 347 frequency channels spanning the 220--440~GHz range. The response function of individual filters follows a Lorentzian profile with a peak coupling efficiency of 13.6\% (based on the laboratory measurements of the first \textsc{Deshima} 2.0 chip). Note that this is a conservative estimate; further sensitivity improvements are expected from the optimization of the chip manufacturing process. We assume a precipitable water vapour (pwv) of 0.6~mm, source elevation of 40 degrees, and a total on-source time of 3~hr. Even with such a short integration, we expect robust detections of multiple CO lines and [CI](2--1). Moreover, we will cover the potentially bright H$_2$O and H$_2$O$^+$ lines.

Finally, we note that due to the relatively low spectral resolution (R$\sim$500), several emission lines might blend together: particularly CO(7--6) and [CI](2--1) (rest-frame frequency separation $\Delta f_0=2.69$~GHz, $\Delta v \sim$ 1000 km/s) and CO(9--8) and OH$^+$ ($\Delta f_0=3.86$~GHz, $\Delta v \sim$ 1100 km/s). The latter pair is particularly susceptible to blending as OH$^+$ often traces gas that is out- or inflowing at high velocities and might be seen in absorption\cite{Riechers2021, Butler2021}; consequently, the CO(9--8) flux measured from $R\sim500$ spectra might be significantly over/underestimated.

\section{Conclusions}

We have presented the high-redshift extragalactic science case for the \textsc{Deshima}~2.0 integrated superconducting spectrometer, which will be mounted at the 10-m ASTE telescope in 2022. Thanks to its combination of an octave-wide bandwidth, access to high frequencies and competitive sensitivity, \textsc{Deshima}~2.0 will allow science-grade observations of high-redshift galaxies. The first integrated chip has been manufactured and tested in the lab. The current performance is already comparable with the Z-Spec grating spectrometer, and after further improvements, should be competitive with APEX. 

In the upcoming Science Verification Campaign, we expect to: (1) measure redshifts for $\sim$30 \textit{Herschel}-selected galaxies at $z\sim1$ and $z\geq4$; (2) obtain multi-line spectroscopy of 5--10 strongly lensed DSFGs from the Planck and SPT samples to study the physical conditions in these extreme sources. These figures are conservative; with further sensitivity improvements, the campaign can be expanded significantly.
These \textsc{Deshima}~2.0 observations will pave the way for future large-scale spectroscopic campaigns with ultra-wideband, multi-pixel MKID spectrometers\cite{Geach2019}. 

\vfill

\begin{acknowledgements}
M.\,R. is supported by the NWO Veni project ``Under the lens'' (VI.Veni.202.225). J.\,B. is supported by the European Research Council ERC (ERC-CoG-2014 - Proposal \#648135 MOSAIC). Y.\,T. and T.\,B. are supported by NAOJ ALMA Scientific Research (grant No. 2018-09B). T.\,T. was supported by MEXT Leading Initiative for Excellent Young Researchers (grant No. JPMXS0320200188). The ASTE telescope is operated by the National Astronomical Observatory of Japan (NAOJ).
\end{acknowledgements}

\pagebreak

\bibliographystyle{./unsrt_truncate.bst}
\bibliography{references} 

\end{document}